\documentstyle[12pt]{article}
\def\etal{et al.\ }
\def\lsim{\stackrel{<}{{}_\sim}}
\def\gsim{\stackrel{>}{{}_\sim}}

\begin{document}

\baselineskip 18pt

\newcommand{\sheptitle}
{Chargino Production at LEP2 in a Supergravity Model}

\newcommand{\shepauthor}
{Marco A. D\'\i az and Steve F. King }

\newcommand{\shepaddress}
{Physics Department, University of Southampton\\
Southampton, SO17 1BJ, U.K.}

\newcommand{\shepabstract}
{In the framework of a particular supergravity model
which provides a natural solution to the $\mu$--problem
we show how the discovery of a chargino at LEP2
and the measurement of its mass and production cross--section,
together with the measurement of the mass of the lightest  
neutralino, would determine the entire Higgs and SUSY spectrum.
We give detailed predictions for the Higgs and SUSY spectrum
as a function of the chargino production cross--section,
for constant values of the lightest chargino and gluino masses.}

\begin{titlepage}
\begin{flushright}
SHEP 95/30\\
hep-ph/9601230\\
January 1996\\
\end{flushright}
\vspace{.4in}
\begin{center}
{\large{\bf \sheptitle}}
\bigskip \\ \shepauthor \\ \mbox{} \\ {\it \shepaddress} \\ \vspace{.5in}
{\bf Abstract} \bigskip \end{center} \setcounter{page}{0}
\shepabstract
\end{titlepage}

\newpage

Recently LEP1.5 has set a new lower bound on the lightest 
chargino mass of about 65 GeV if 
$m_{\tilde\chi^{\pm}_1}-m_{\tilde\chi^0_1}\gsim 10$ GeV 
\cite{LEP15}. In general LEP2 will be
able to bound or discover charginos up to the kinematic
limit of the machine. In this paper we explore the possible
consequences of chargino discovery at LEP2 within the 
framework of a well motivated supergravity model.

It is well known that, with the assumption of a universal
gaugino mass $M_{1/2}$, the chargino
$\tilde{\chi}^{\pm}_i$ ($i=1,2$)
and neutralino $\tilde{\chi}^0_i$ ($i=1 \ldots 4$) masses
and mixing angles only depend on three unknown parameters:
the gluino mass $m_{\tilde{g}}$, $\mu$ and $\tan \beta$ 
\cite{GunionHaber}. In a recent paper \cite{dk} we have shown how
the discovery of the lightest chargino at LEP2
and the measurement of its mass, $m_{\tilde{\chi}^{\pm}_1}$, 
and production cross-section,
$\sigma (e^+e^-\rightarrow \tilde{\chi}^+_1 \tilde{\chi}^-_1)$,
together with the measurement of the mass of the lightest  
neutralino, $m_{\tilde{\chi}^{0}_1}$, will enable 
the basic parameters $m_{\tilde{g}}$, $\mu$ and $\tan \beta$
to be determined, up to certain ambiguities (see also ref. 
\cite{Feng}).

In the present paper we shall extend the above analysis from the
gaugino sector of the Minimal Supersymmetric Standard Model 
(MSSM) \cite{MSSMrep} to the entire supersymmetric (SUSY)
and Higgs spectrum. However, whereas the gaugino sector
is completely specified by three parameters, the remaining spectrum
depends on very many parameters and without some simplifying principle
it is impossible to make progress. Therefore in the present paper
we shall explore the consequences of a specific supergravity 
(SUGRA) model
which has sufficient predictive power to enable the entire Higgs 
and SUSY spectrum to be deduced from just the LEP2 measurements
$m_{\tilde{\chi}^{\pm}_1}$, 
$\sigma (e^+e^-\rightarrow \tilde{\chi}^+_1 \tilde{\chi}^-_1)$,
and $m_{\tilde{\chi}^{0}_1}$ -- a result which underlines
both the importance of LEP2 and the power of supergravity.

The phenomenologically simplest SUGRA models typically 
involve universal soft parameters (at the unification scale
\footnote{We shall take apply these boundary conditions at 
the gauge coupling unification scale $M_X$, neglecting
any effects due to running between the Planck scale and $M_X$.}):
$m_0$, $M_{1/2}$, $A$, $B$ in the usual notation corresponding to the
universal scalar mass, gaugino mass, trilinear dimensionful coupling and 
$B\mu H_1H_2$ term, respectively. Thus the squark and slepton soft masses
are proportional to unit matrices in flavor space, and trilinear
couplings are proportional to Yukawa matrices at the unification scale.
Specific SUGRA models may involve further relationships between the
soft parameters, for example the so-called minimal SUGRA model
predicts that $B=A-m_0$ \cite{SUGRAI}. However this model involves
an unnaturally small dimensional $\mu$ parameter appearing in the
superpotential -- the $\mu$ problem. Recently there have been
several alternative mechanisms proposed to solve the $\mu$ problem
\cite{SUGRAII} and it is a common tendency of such models
to predict $B=2m_0$, although it is not clear why such different
theories should lead to the same boundary condition.
Therefore in the present paper we shall focus on SUGRA models
which predict $B=2m_0$ which have a stronger theoretical motivation.

According to the above discussion, the 5 independent parameters at
$M_X$ are: $B=2m_0$, $M_{1/2}$, $A$, $\mu$, $h_{t0}$,
where $h_{t0}$ is the top quark Yukawa coupling at $M_X$.
We shall require radiative electroweak symmetry breaking, and impose 
the usual Higgs minimisation conditions at low energy. 
We consider all the supersymmetric mass parameters to be smaller
than $M_{SUSY}=$1 TeV. The order of magnitude of 
this scale emerges naturally when the model 
is embedded into a GUT \cite{GUT}. In addition, to break radiatively the 
electroweak symmetry without fine--tunning the initial parameters
$M_{SUSY}$ cannot be too large \cite{BdeCACfine}.
Our 4 input parameters are chosen to be:
$m_t$, $m_{\tilde{\chi}^{\pm}_1}$, $\mu$ and $m_{\tilde{g}}$ which 
are sufficient to specify the 5 independent parameters
at $M_X$, given the requirement of correct electroweak symmetry breaking.
The idea behind this choice of input parameters is that the top
quark mass is measured at the Tevatron, and the chargino mass
may be measured at LEP2. This only leaves the parameters
$\mu$ and $m_{\tilde{g}}$ which can in principle be determined at LEP2
from a measurement of 
$\sigma (e^+e^-\rightarrow \tilde{\chi}^+_1 \tilde{\chi}^-_1)$ and
$m_{\tilde{\chi}^{0}_1}$, as discussed in
our previous analysis \cite{dk} except that now the electron
sneutrino contribution to the cross-section will be taken into account.
The main difference is of course that now these LEP2 measurements
will serve to
determine the {\em entire} Higgs and SUSY spectrum, not just the
gaugino sector.

Our detailed procedure is to first fix values of
top quark mass, chargino mass and gluino mass. For a given choice
of $\mu$, knowledge of $m_{\tilde{\chi}^{\pm}_1}$ 
and $m_{\tilde{g}}$ enables a
determination of $\tan \beta$. With $m_t$ and $\tan \beta$
specified we have a determination of $h_t$ (at low energy)
and hence $h_{t0}$ (at high energy).
With $h_{t0}$ known, we choose values of $m_0$ 
and $A$ and run all the parameters
down to low energy (the RG equations do not depend on $B$). 
We do not take into account threshold corrections.
The tree--level minimisation condition on the 
Higgs masses at low energy
\footnote{Each of these terms is 
equal to $\frac{1}{2} m_A^2(1-\cos^2 2\beta)$,
where $m_A$ is the CP-odd scalar mass.} 
\begin{equation}
(m_1^2+\mu^2+\frac{1}{2} M_Z^2\cos 2\beta)(1+\cos 2\beta)=
(m_2^2+\mu^2-\frac{1}{2} M_Z^2\cos 2\beta)(1-\cos 2\beta)
\label{eq:minim}
\end{equation}
where $m_1^2$ and $m_2^2$ are the soft SUSY breaking Higgs masses,
will not in general be satisfied, and so we vary $m_0$ until 
it is. Sometimes there will be no solution for any value of $m_0^2>0$,
and this condition has a big 
effect in reducing the allowed parameter space.
Eq.~(\ref{eq:minim})\ describes the minimization of the Higgs 
potential when the one-loop contributions to the effective potential 
are neglected. 
Having consistently determined $m_0$ we then 
find the low energy value of $B$ using
$B\mu=\frac{1}{2}m_A^2\sin 2\beta$, and run it up to
find the high energy value of $B$. In general the condition
$B=2m_0$ will not be satisfied, and we iterate the procedure
for different values of $A$ until this condition is met, which
effectively serves to determine $A$. If we have chosen a suitable
value of $\mu$ we will then have a successful data point from
which the entire Higgs and SUSY spectrum may be calculated,
and $\sigma (e^+e^-\rightarrow \tilde{\chi}^+_1 \tilde{\chi}^-_1)$
computed. 
We note that there is no ambiguity in the 
sign of the supersymmetric Higgs
mass parameter $\mu$, as opposed to global supersymetry
analyzed in ref.~\cite{dk}, because the high energy relation 
$B=2m_0$ together with the low energy constraint
$m_A^2=2B\mu/\sin(2\beta)>0$ implies that only $\mu>0$
solutions are satisfactory \cite{SUGRAII} (see also \cite{BdeCyC}). 

Using the above procedure, contours in the
$\sigma (e^+e^-\rightarrow \tilde{\chi}^+_1 \tilde{\chi}^-_1)$ -- 
$m_{\tilde{\chi}^{0}_1}$ plane for a fixed values
of $m_t$, $m_{\tilde{\chi}^{\pm}_1}$ and $m_{\tilde{g}}$
are produced by varying $\mu$ over its successful range.
These contours are shown in Figure 1.
We see four groups of curves corresponding
to the choices $m_{\tilde\chi^{\pm}_1}=60$, 70, 80, and 90 GeV.
We have taken $m_t=176$ GeV and $\alpha_s=0.116$.
We do not find acceptable solutions for $m_{\tilde\chi^{\pm}_1}=50$, 
consistent with the LEP1 experimental lower bound on the chargino mass
$m_{\tilde\chi^{\pm}_1}>45$ GeV \cite{ExpRep,selchaexp,chaexp}.
For each value of the chargino mass, the different curves are 
labeled by the value of the gluino mass. It is clear from the figure 
that heavier charginos produce smaller total cross sections and, at
the same time, they are associated with heavier neutralino 1,
$\tilde\chi^0_1$.
This neutralino is the lightest supersymmetric particle, or LSP.
Since the LSP should be neutral, some curves for the case
$m_{\tilde\chi^{\pm}_1}=90$ GeV are truncated at high values of $\tan\beta$
because beyond that point the $\tilde\tau^{\pm}_1$ becomes lighter than
$\tilde\chi^0_1$. It can also be appreciated from Fig.~1 that larger 
$m_{\tilde g}$ produce, in general, smaller cross sections. The reason 
here is that the sneutrino becomes lighter at larger gluino masses,
and since the sneutrino contribution to the total cross section
is negative, the total cross section decreases.
Note that contours in Fig.~1 are truncated at the upper end (large
cross--section end) where $\tan\beta$ is small, and at the lower
end (small cross--section end) where $\tan\beta$ is large, for 
reasons which will become apparent.
Fig.~1 clearly demonstrates that for a given $m_t$, LEP2 measurements
of $m_{\tilde\chi_1^{\pm}}$, 
$\sigma (e^+e^-\rightarrow \tilde{\chi}^+_1 \tilde{\chi}^-_1)$
and $m_{\tilde\chi^0_1}$ are sufficient to completely specify
all the parameters in the theory.

To emphasise that
all the basic parameters of the theory, and with them,
the whole Higgs and SUSY spectrum, are determined for each point of each
contour in Fig.~1, we shall present a series of contours with
$\sigma (e^+e^-\rightarrow \tilde{\chi}^+_1 \tilde{\chi}^-_1)$
along the vertical axis and some other determined quantity along the 
horizontal axis. In Figure 2 we show the set of $m_{\tilde\chi_1^{\pm}}$,
$m_{\tilde g}$ contours corresponding to Fig.~1 but with the 
basic parameters (a) $m_0$, (b) $A$, (c) $\tan \beta$ and (d) $\mu$ 
along the horizontal axes. For a given chargino mass, the gluino 
masses we have taken are bounded from above because 
gluinos heavier than some value 
produce solutions with $m^2_0<0$ [Fig.~2(a)], and from below
because gluinos lighter than some value
need values of $A$ larger than 1 TeV [Fig.~2(b)]. We do not consider
constraints form charge and color breaking \cite{Casas}. 
The parameter $\tan\beta$ is plotted in Fig.~2(c),
and the most noticeable feature is that many curves are 
truncated at $\tan\beta\approx2$. The reason is that we are close
to the fixed point of the top quark Yukawa coupling, and smaller
values of $\tan\beta$ makes this coupling diverge at scales
smaller than the unification scale. In the case of the 
lightest gluino choice, the fixed point of $h_t$ is not reached,
and the curve is truncated at higher values of $\tan\beta$
because the parameter $A$ becomes larger than 1 TeV.
For a fixed value of $m_{\chi^{\pm}_1}$ and $m_{\tilde g}$, 
the parameter $\mu$ is determined by the value of $\tan\beta$
and it is plotted in Fig.~2(d). Typically, the smaller the
chargino mass is, the smaller the parameter $\mu$ is. Nevertheless,
it never reaches values smaller than 150 GeV.

In Figure 3 we show the contours corresponding
to Figure 1 but with (a) the second lightest neutralino mass,
(b) the sneutrino mass, (c) the lightest charged slepton mass
and (d) the lightest up-type squark mass along the horizontal axis.
In Fig.~3(a) we see the production cross section as a function
of $m_{\chi^0_2}$. This mass is strongly correlated with the
lightest chargino mass, satisfies $m_{\chi^0_2}\gsim m_{\chi^{\pm}_1}$,
and receives small increases as we increase the gluino mass. 
Similarly to the lightest neutralino case, the groups of curves
corresponding to different values of $m_{\chi^{\pm}_1}$ are
well differentiated, implying that we will have a very good
idea of the mass of this particle even if we have large experimental 
errors on the LEP measurements mentioned before. The sneutrino mass
is presented in Fig.~3(b). Note that it is the electron sneutrino
that is of interest to us because it contributes to the
chargino production cross section. Nevertheless, the three
sneutrino flavors are practically degenerated in mass. The 
sneutrino mass always satisfies the experimental constraint
$m_{\tilde\nu}>41.8$ GeV \cite{ExpRep,sneuExp}
and is represented in the figure by a
vertical dotted line. In can be appreciated that the sneutrino
contribution to the total chargino production cross section is
more important when charginos are light, and that it decouples
as $m_{\tilde\nu}$ increases. In Fig.~3(c) we plot the lightest
charged slepton mass, which in all cases is $\tilde\tau^{\pm}_1$. 
The experimental LEP1 constraint $m_{\tilde\tau^{\pm}_1}>45$ GeV
\cite{ExpRep,selchaexp,selexp}
restricts the allowed parameter space, and consequently some of the 
curves (the ones with lighter gluino) are truncated at large 
$\tan\beta$. The reason for that lies in the left--right mixing of the
mass matrix, because it is proportional to $m_{\tau}\mu\tan\beta$, and
large values of $\tan\beta$ produce a large mass splitting between
$\tau^{\pm}_1$ and $\tau^{\pm}_2$. The lightest of the up--type
squarks is plotted in Fig.~3(d), which is 
predominantly the lightest stop, $\tilde t_1$, 
with a small component of scharm (in all cases smaller than a
percent). At small values of the universal scalar mass parameter 
$m_0$, where the total cross section is small, the mass of the 
lightest up--type squark increases with this parameter $m_0$. 
Nevertheless, for large values of $m_0$ the trilinear mass parameter 
$A$ also is large, producing a large top squark mass mixing and
consequently, a lighter $\tilde t_1$. This makes the curves in
Fig.~3(d) to turn towards the small values of the lightest
up--type squark as the total cross section increases. Lighter
$\tilde t_1$ are obtained when small chargino masses are considered,
a combination that produces large corrections to the 
$Z\rightarrow b\bar b$ decay. Nevertheless, we do not find top
squarks lighter than about 180 GeV, claimed to be necessary to 
explain the discrepancy between theory and experiment
(see for example \cite{Zbb}).

The Higgs sector of the MSSM \cite{HHG} is completely specified at
tree level by the CP-odd Higgs mass $m_A$ and $\tan\beta$. 
Nevertheless, a strong dependence on $m_t$ and $m_{\tilde t}$
is introduced through radiative corrections to the charged Higgs
mass \cite{ChargedH,chaneu} and to the lightest neutral
CP-even Higgs mass \cite{chaneu,neutral}.
In Figure 4 we show the corresponding contours of Fig.~1 with (a) the
lightest CP-even Higgs boson mass, (b) the CP-odd Higgs boson mass,
(c) the charged Higgs boson mass and (d) the value of 
$\cos (\beta - \alpha)$ along the horizontal axes. In this scenario, 
the lightest Higgs mass plotted in Fig.~4(a) 
is always smaller than 103 GeV, with obvious relevance for LEP2
\cite{SMhiggs}. The higher
values of $m_h$ are obtained when $\tan\beta$ is large, where
the tree level contribution to $m_h$ is maximum. On the other
hand, at $\tan\beta\sim2$ the lightest Higgs mass is minimum,
with a lower bound of about 80 GeV. In the calculation of $m_h$
we include the exact one-loop radiative corrections from top
and bottom quarks and squarks and leading logarithms from the
rest of the particles, working in an on-shell scheme where the 
parameter $\tan\beta$ is defined through the
$A\tau^+\tau^-$ coupling \cite{diazi}.
We also include the dominant two-loop QCD
corrections \cite{RalfHo} and we sum all the leading and 
next-to-leading logarithms with a RGE technique. 
The mass of the CP-odd Higgs $m_A$ is plotted in Fig.~4(b).
It is obtained from eq.~(\ref{eq:minim})\ which comes from
the minimization of the Higgs potential. This particle is in
all cases heavier than about 130 GeV, what makes it difficult
to be observed at LEP2. The charged Higgs mass, plotted
in Fig.~4(c), is strongly
correlated to the value of $m_A$, because at tree level the
relation $m_{H^{\pm}}^2=m_W^2+m_A^2$ holds. We include the one-loop
radiative correction to this mass, nevertheless, in the region
of parameter space considered here, the correction is smaller
than $\sim 2$ GeV in all cases. Finally, in Fig.~4(d) we plot
the parameter $-\cos(\beta-\alpha)$ which controls the $ZZH$
coupling. One-loop corrections to
this angle are already taken into account \cite{marcodpf}.
The fact that this parameter
remains small implies that the heavy Higgs boson $H$ is weakly
coupled to the $Z$--boson. At the same time the lightest
Higgs $h$, with a $ZZh$ coupling proportional to $\sin(\beta-\alpha)$,
has couplings that approach the corresponding Standard Model (SM) 
Higgs boson couplings. Nevertheless, it is worth mentioning that,
if we consider the SM with no new physics below $\sim10^{10}$ GeV and 
the MSSM with $M_{SUSY}\lsim1$ TeV, the allowed values of $m_{H_{SM}}$ 
are always greater than $m_h$ providing the top quark is sufficiently
heavy \cite{MSSMorSM}. 

Finally we return to our original question: what can we learn from 
this SUGRA model with the detection of charginos at LEP2?
The answer is summarised in Fig.~1. In this figure a
group of curves corresponding to a particular value
of $m_{\chi^{\pm}_1}$ are well differentiated from a group
corresponding to a different value. This permits us to check the
validity of the model even if the experimental errors are so large
that it is not possible to precisely differentiate the curves labeled
by the value of the gluino 
mass. Assuming that precise measurements of the total cross
section $\sigma(e^+e^-\rightarrow\tilde{\chi}^+_1\tilde{\chi}^-_1)$,
the chargino mass $m_{\tilde\chi^{\pm}_1}$, and the lightest
neutralino mass are available, it is possible to predict the
value of $m_{\tilde g}$, and from Figs.~2--4 the rest of the 
parameters of the model and the masses of the physical particles
can be predicted within this particular well motivated SUGRA
model. For example a firm prediction of this model is that if 
LEP2 discovers a chargino then the lightest CP-even Higgs 
boson must be lighter than 103 GeV and have SM--like couplings.

\vspace{0.25in}

\section*{Acknowledgements}

One of us (MAD) has benefited from discussions with B. de Carlos 
on different aspects of this problem and is thankful to A. Casas for 
clarifying the origin of the relation $B=2m_0$.

\newpage

\noindent
{\bf \large Figure Captions:}

\noindent
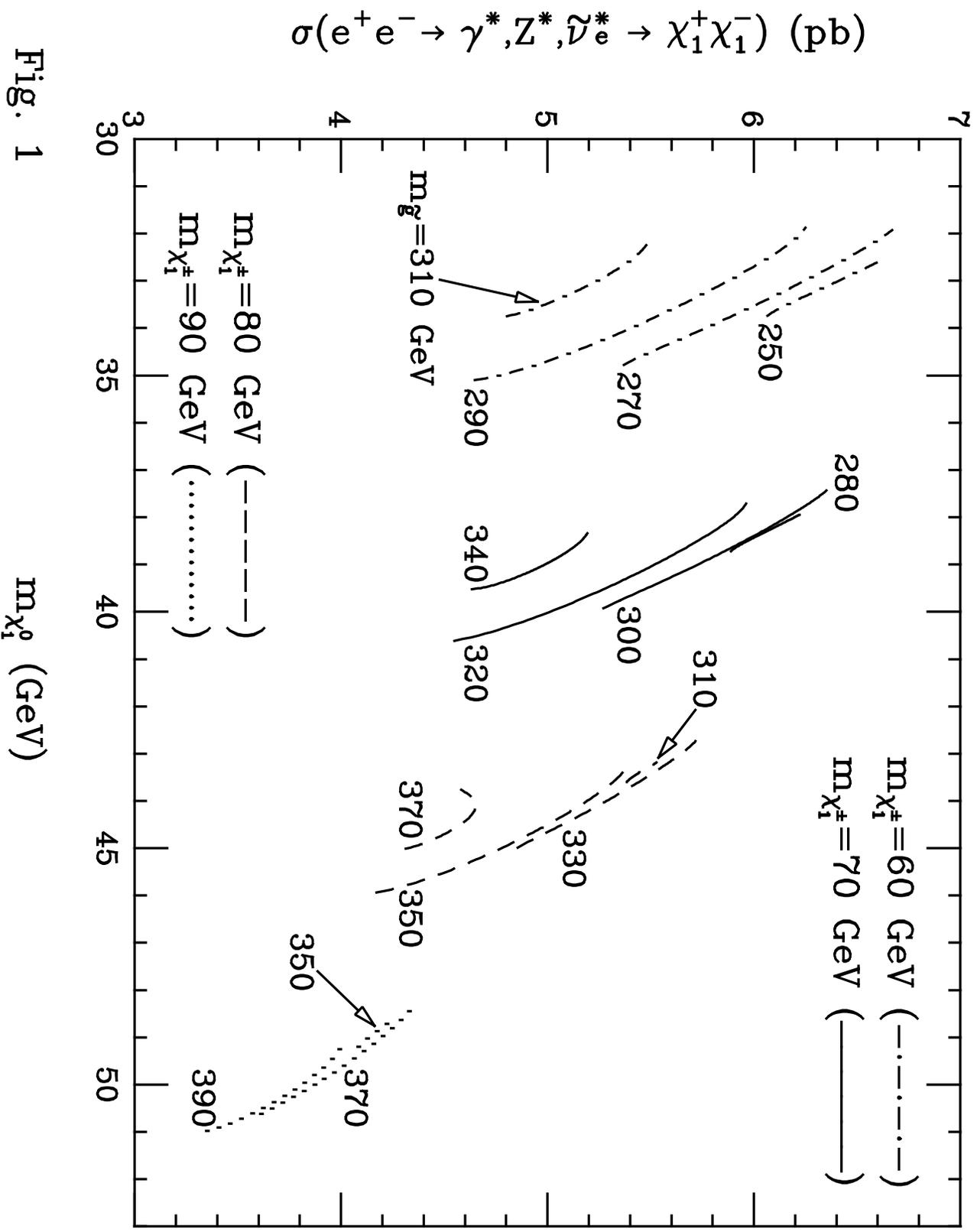
{\bf Fig. 1.} 
Total cross section of chargino pair production from
$e^+e^-$ annihilation as a function of the mass of the 
lightest neutralino for constant values of the lightest
chargino and the gluino masses. Four groups of curves are
shown corresponding to $m_{\tilde\chi_1^{\pm}}=60$ GeV (dotdash),
70 GeV (solid), 80 GeV (dashes), and 90 GeV (dots), with
each line labelled by the gluino mass in GeV.
\\
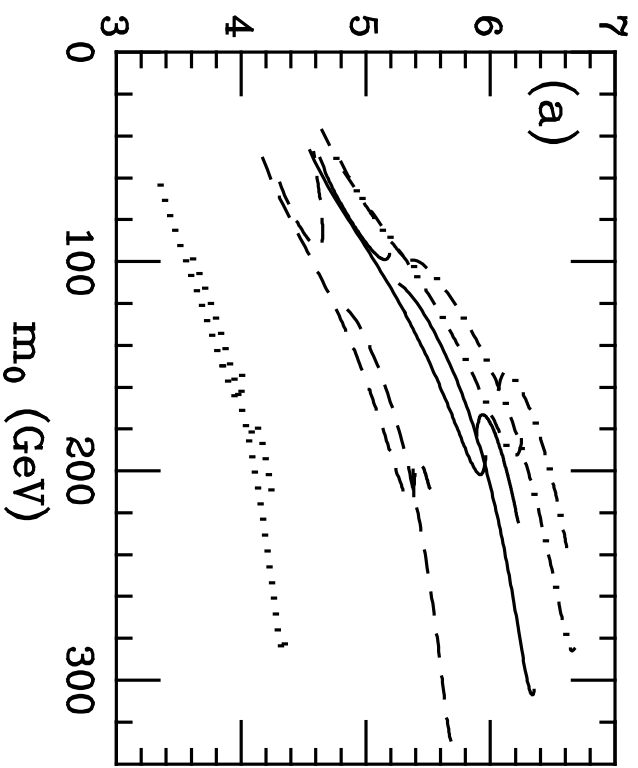
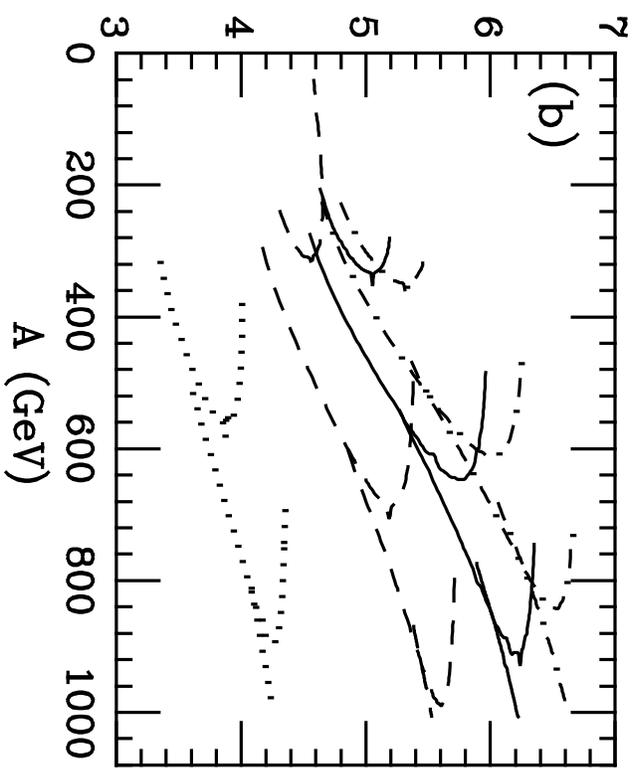
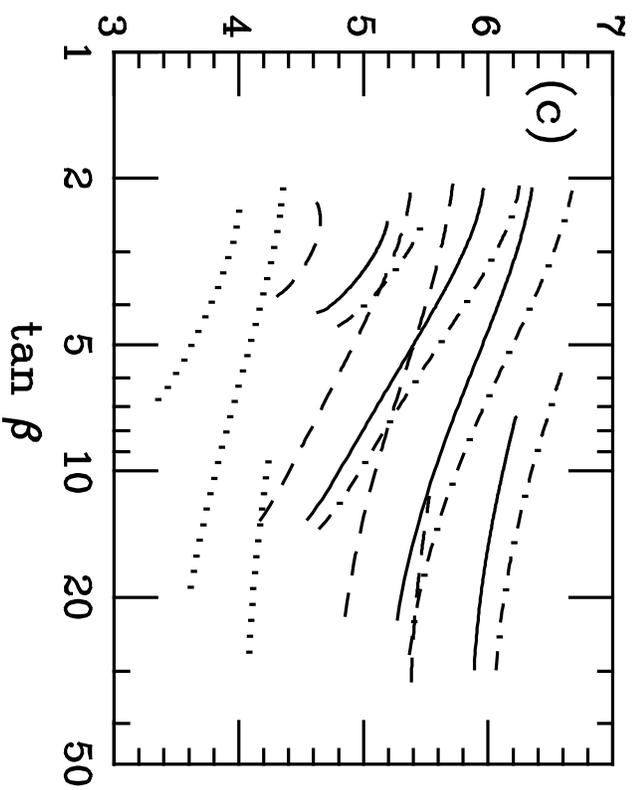
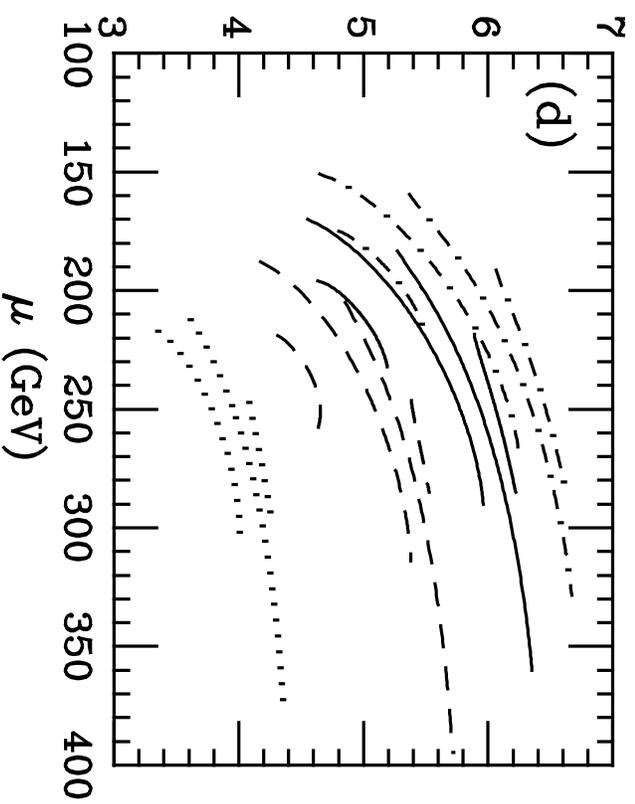
{\bf Fig. 2.}
Total chargino pair production cross section as a function of
four different parameters that characterize the supergravity
model. (a) The universal scalar mass $m_0$, (b) the universal 
trilinear coupling $A$, (c) the ratio of the two Higgs vacuum 
expectation values $\tan\beta$, and (d) the Higgs mass parameter
$\mu$. There is a one--to--one correspondence between the lines
in this Figure and those in Figure 1. The line styles for the
chargino masses are as in Figure 1. For each chargino mass the
gluino masses also correspond to those in Figure 1. Although
we have not labelled the gluino mass, it should be possible to
distinguish which line corresponds to which gluino mass by
comparing the total cross--section for chargino production
plotted here to that plotted in Figure 1.
\\
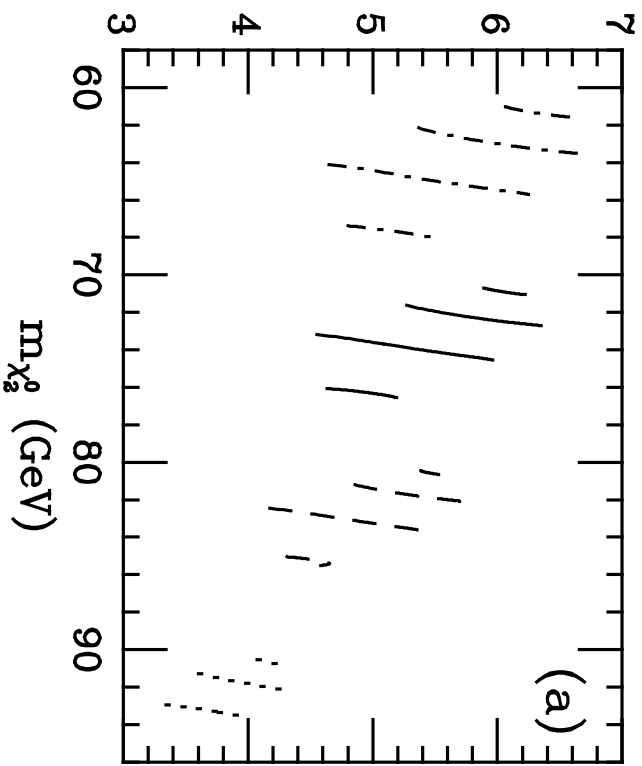
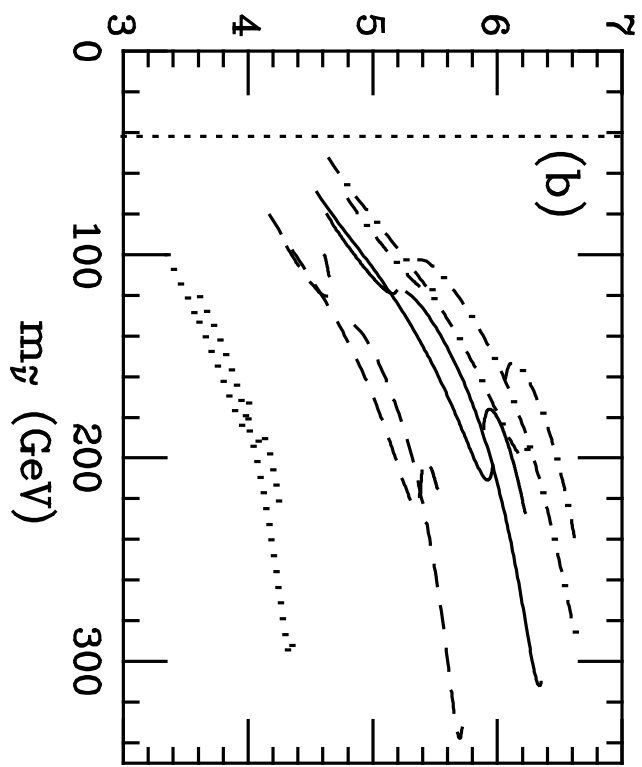
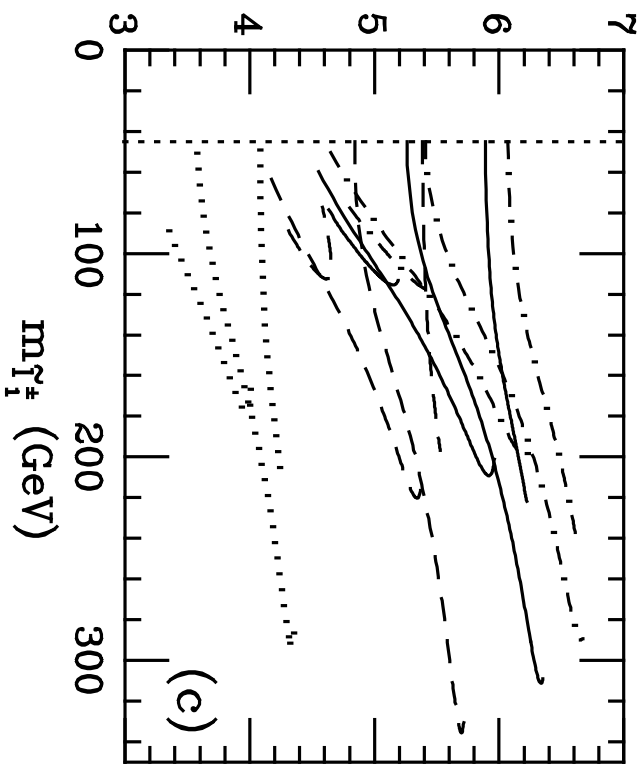
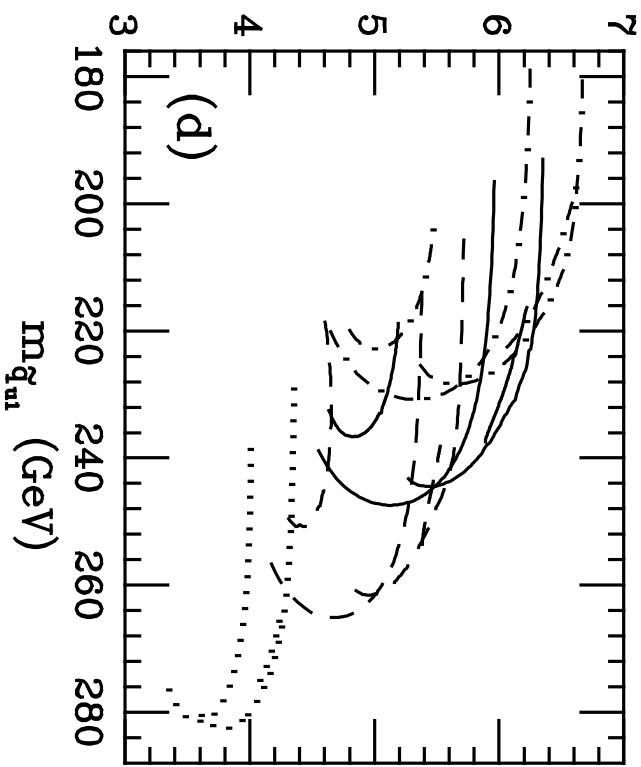
{\bf Fig. 3.}
Total chargino pair production cross section as a function
of the mass of (a) the second lightest neutralino, (b) the
sneutrino, (c) the lightest charged slepton, and (d) the 
lightest up--type squark. 
There is a one--to--one correspondence between the lines
in this Figure and those in Figure 1. The line styles for the
chargino masses are as in Figure 1. For each chargino mass the
gluino masses also correspond to those in Figure 1. Although
we have not labelled the gluino mass, it should be possible to
distinguish which line corresponds to which gluino mass by
comparing the total cross--section for chargino production
plotted here to that plotted in Figure 1.
\\
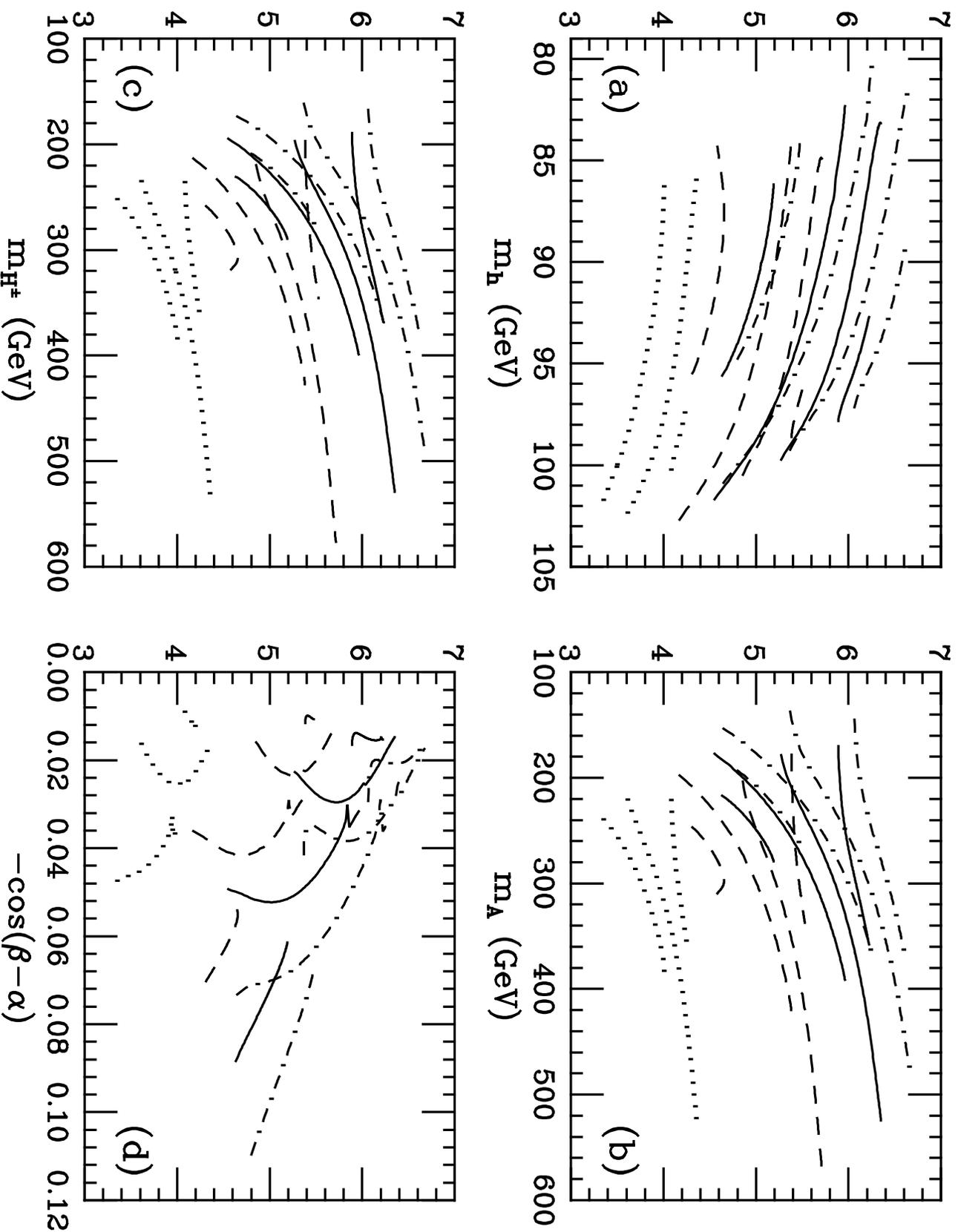
{\bf Fig. 4.}
Total chargino pair production cross section as a function of 
four different parameters in the Higgs sector. (a) The lightest
Higgs mass, (b) the CP-odd Higgs mass, (c) the charged Higgs 
mass, and (d) the parameter $-\cos(\beta-\alpha)$ whose 
magnitude is the coupling of the heavy CP-even Higgs to a pair
of $Z$--bosons. There is a one--to--one correspondence between the 
lines in this Figure and those in Figure 1. The line styles for the
chargino masses are as in Figure 1. For each chargino mass the
gluino masses also correspond to those in Figure 1. Although
we have not labelled the gluino mass, it should be possible to
distinguish which line corresponds to which gluino mass by
comparing the total cross--section for chargino production
plotted here to that plotted in Figure 1. 
\\

\end{document}